\documentclass[aps,superscriptaddress,showkeys,showpacs]{revtex4-1}
\usepackage{graphicx}
\usepackage{dcolumn}
\usepackage{bm}

\usepackage{xcolor}
\usepackage{lineno}
\usepackage[capitalize]{cleveref}

\begin{document}

\title{Information and Configurational Entropy in Glassy Systems}

\author{Ittai Fraenkel}
\affiliation {Department of Physics, Technion-IIT, {32000} Haifa, Israel}
\author{Jorge Kurchan} 
\affiliation {Laboratoire de Physique de l'\'Ecole Normale Supérieure, ENS, Université PSL, CNRS, Sorbonne Université, Université de Paris, F-75005 Paris, France}
\author{Dov Levine}
\affiliation {Department of Physics, Technion - IIT, {32000} Haifa, Israel}




\begin{abstract}


It is often stated that if one is presented with a snapshot of
the positions of the molecules of a glass and one of a liquid, one is unable to tell the difference.
Here we argue instead that given several such snapshots taken over a time-interval,
{\it even without specifying the times},
there is a definite procedure to assess precisely the level of glassiness: it suffices
to concatenate the snapshots side-by-side, and to subject the joint picture to a lossless compression protocol. We argue that the size of the compressed file
yields a direct and unambiguous measure of the `vibrational' and `configurational' entropies, and may be used to study the associated glass length scale in or out of equilibrium through the size and frequency of the repeated motifs essential to the compression,
 a quantity that would diverge at a putative glass transition.
\end{abstract}

\maketitle


As a liquid approaches the glass transition, motion gradually slows. Its molecules perform rapid jittery motion in place - just as in a crystal - but in addition there remain spatial density fluctuations which evolve on a slow timescale $\tau_\alpha$ (the `$\alpha$ timescale'), of the order of the inverse of the viscosity. This timescale grows abruptly as temperature is lowered, density is increased, or, more frequently, as the system `ages' after having been quenched in a low temperature bath. 

Considering an equilibrium supercooled liquid\cite{SuperCooledLiquidsForPedestrians}, one may split the entropy into two parts: the first, a `vibrational entropy' due to the rapid motion of particles, and the second, a `configurational entropy', or `complexity', resulting from the exploration of phase-space over times larger than the $\alpha$ scale. Schematically speaking, the vibrational entropy measures the motion of molecules about a (temporarily) rigid backbone, while the configurational entropy reflects the number of possible backbones, sampled over a much longer time scale. Over seventy years ago, Kauzmann \cite{kauzmann1948nature} proposed that, for glasses, the first contribution could be estimated to be of the order of the entropy of a crystal, leaving the configurational entropy to make up the rest \footnote{Because we shall work in regimes (liquid in equilibrium or aging) in which the `backbones' are only temporary,
the definition of complexity we discuss here is in the thermodynamic but limit inevitably based on a timescale, and thus makes sense 
only in  relation to a separation of fast (`vibrational') and slow (`alpha') timescales. This is unlike a construction applied to
the equilibrium Gibbs measure, see  References \cite{holler2020one,read2022complexity,newman2022metastates}, which apply to infinitely long equilibration times.}.

Mathematically speaking, expressing by $x$ the vibrational configurations about a given backbone $y$, and the associated joint and conditional probabilities as $P(x,y), P(x|y)$ respectively, we have that the total entropy is: 
\begin{eqnarray} & S_{tot} = &\sum_{x,y} P(x,y) \log P(x,y) \nonumber\\ &=& \sum_y P(y) \sum_{x} P(x|y) \Big( \log P(x|y)+ \log P(y) \Big) \nonumber\\ &=& \sum_y P(y) S_{vib}(y) + \sum_y P(y) \log P(y) \\ &=& S_{vib} + S_{config}\nonumber \end{eqnarray}

The configurational entropy thus defined decreases with temperature, and a straightforward extrapolation determines a `Kauzmann' temperature where it would become zero \cite{kauzmann1948nature}. This temperature corresponds to a putative `ideal glass transition' which is unreachable experimentally in a step-by-step equilibrium protocol.
The dynamic consequences of this scenario were explored by Gibbs and Di Marzio \cite{gibbs1958nature}, and by Adam and Gibbs \cite{adam1965temperature}, who proposed phenomenological models to illustrate how configurational entropy decrease controls the slowing down of the $\alpha$-scale. The Adam-Gibbs-Di Marzio picture has been discussed - and criticized - extensively in the literature, but one fact remains certain:  simple nucleation arguments \cite{kurchan2010order,berthier2011overview}, and rigorous treatments \cite{van1984statistical,read2022complexity} show that in the presence of finite equilibrium configurational entropy, there can be no stable amorphous states. These can only exist if and when the configurational entropy reaches zero (that is, when it is subextensive in system size).

 Several methods have been employed to quantify configurational and vibrational entropies of glassy systems; these are implemented and discussed in detail in Reference \cite{berthier2017configurational}. One rough procedure uses thermodynamic integration to calculate the total entropy of a system, and estimates the vibrational
 entropy to be that of a crystal at the same temperature. The 
 configurational entropy is then evaluated as the difference of these values.
 An improved version of this procedure evaluates the vibrational entropy using a Frenkel-Ladd (F-L) integration \cite{frenkel1984new}: the particles are connected by springs to their positions in a reference equilibrium configuration, and then annealed over the spring stiffness. Although more quantitative, it is not clear if it takes into account all fast processes, such as particle exchanges, and it cannot be applied to non-particulate systems with glassy dynamics, where it is not well-defined. Another method, that of the Franz-Parisi potential \cite{franz1998effective}, studies the overlap $Q$ with a reference configuration: an equilibrium configuration at temperature $T$ is frozen and used as a substrate to which a second unfrozen configuration at the same temperature is coupled \footnote{The Franz-Parisi method relies on the reference system being in equilibrium, otherwise the connection to the relevant theory is lost.}. The free energy of the system at fixed $Q$ is expected, in the context of the `Random First Order' theory \cite{kirkpatrick1987p,kirkpatrick1989scaling,lubchenko2006theory}, to show a linear dependence on $Q$ (corresponding to a Maxwell construction) \cite{ConfigEntropyCalcBerthier}. The configurational entropy is then obtained in similar manner to the estimation of the latent heat in a first order transition, with the important difference that there is no obvious way to visualise the competing phases, as is possible for a typical first order transition.
 The last method involves the point-to-set length of the system \cite{bouchaud2004adam}, which may be thought of, algorithmically, as follows: The system under study is equilibrated, following which it is frozen. Next, a convex portion is melted and allowed to relax. This newly relaxed cluster is then compared with the original cluster which was present before melting. If the diameter
of the cavity is small enough, they will coincide. The maximal cavity diameter at which this happens defines the point-to-set length, which is then connected to the entropy via a phenomenological relation.

Although compelling, these methods do not afford a direct understanding of the configurational entropy of a system, without invoking any specific theory.
In this Letter, we propose an unambiguous procedure to define and measure the configurational and vibrational entropies that is practical, general, and transparent. Of course, in order for these quantities to be well-defined, the systems we study should have ``fast" and ``slow" degrees of freedom - that is, a separation of time scales. Thus, it is particularly, but not uniquely, relevant to glassy systems.

In recent years, physicists' views on entropy have been influenced by ideas from information theory, in particular lossless data compression \cite{cover2012elements}. Data compression algorithms take advantage of patch repetitions in an image or text file in order to generate a shorter file (the ``encoding''), which may be decoded to reproduce the original file exactly. Optimal compression protocols, such as the various Lempel-Ziv \cite{ziv1977universal,ziv1978compression} algorithms, produce (in the asymptotic limit of large data files) the shortest possible encoding, whose average length is the Shannon entropy of the source which generated the data file \cite{shannon2001mathematical}. In equilibrium physics terms, the source is the ensemble of microstates of the system, and the average length of the encodings of compressed microstates is the thermodynamic entropy of the system; this is a succinct way to express the relation
between entropy and information \footnote{As with any classical entropy, one has to identify all continuous quantities up to an error $\epsilon$: densities, and, when necessary, positions, radii and eccentricities of particles. The results then scale logarithmically with the coarse-graining scale $\epsilon$ which drops off in the thermodynamic limit.}.
In practice, even the length of the encoding of a single typical microstate is sufficient to give an excellent estimate of the entropy \cite{martiniani2019quantifying,avinery2019universal}. 

An advantage of this procedure is that it is not limited to equilibrium systems. Indeed, it has been shown to reveal dynamic phase transitions \cite{martiniani2019quantifying} and correlations \cite{martiniani2020correlation} in a variety of non-equilibrium systems, even when the nature of their ordering is not known {\it a priori}. Moreover, incorporation of the time domain \cite{cavagna2020vicsek} has proven fruitful in studies of dynamic models such as the Viscek \cite{vicsek1995novel} flocking model, as well as providing a quantitative measure of time-reversal invariance and entropy production \cite{ro2022model}. 

\begin{figure}[t]
 \centering
 \includegraphics[width=.6\linewidth]{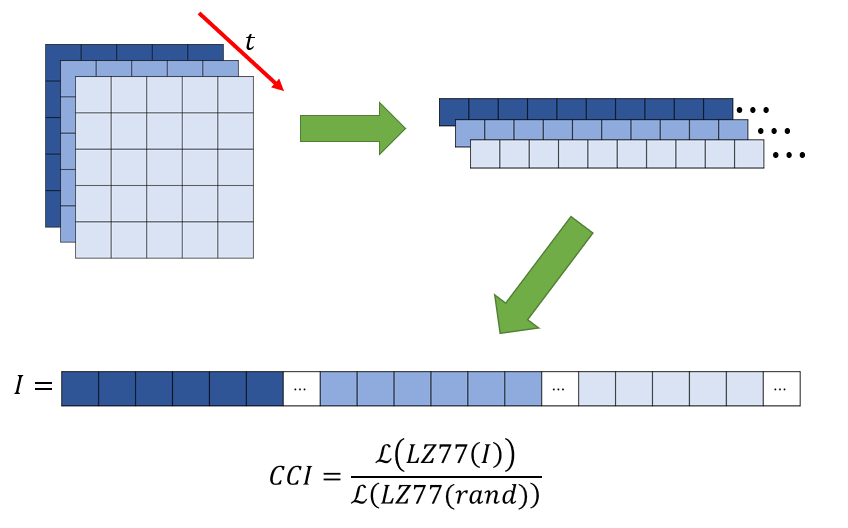}
 \caption{A demonstration of the CCI calculation method. Different snapshots of our system, taken at different times, are projected onto 1D and concatenated. The resultant string \(I\) is then compressed using LZ77 algorithm. The CCI is the length of the compressed string divided by the length a compressed randomized string of the same length.}
 \label{CCI calculation diagram}
\end{figure}

As mentioned above, the present study relies on the large separation of time scales present in glassy systems, which allows us to define vibrational entropy in a natural way by factoring out the (slow) configurational part of the entropy.
To do this, we start with $N$ ($d$-dimensional) snapshots of our system, taken at equal intervals in the time window $(t_w,t_w + \tau)$, where $t_w$ is the waiting time. 
We concatenate these snapshots into a $d+1$ dimensional string (see Fig. \ref{CCI calculation diagram}), which we then compress serially, snapshot by snapshot, following \cite{cavagna2020vicsek}. If $\tau<\tau_\alpha$, the backbone is essentially the same in all the snapshots, allowing for efficient compression: effectively, we need only encode the backbone once, and refer to this encoding in all subsequent encounters. The bulk of the encoding of the concatenated string is used for the vibrations and other fast excitations around the backbone, which are different in each snapshot. In this case, the length of the file encoding the concatenated sequence is:
$$
{\cal L}_{tot} \approx {\cal L}_{backbone}+ \sum_{j=1}^{N}{\cal L}_{vib}^j \;\approx \; {\cal L}_{backbone}+ N{\bar {\cal L}}_{vib}
$$
where ${\cal L}_{vib}^j$ is the length of the file needed to encode the $j^{th}$ configuration, excluding the backbone, and ${\bar {\cal L}}_{vib}$ is its average. Since LZ encoding is asymptotically optimal \cite{cover2012elements}, ${\cal L} \rightarrow S$ in the large system limit, so
${\cal L}_{tot}\rightarrow S_{config}+ N S_{vib}$, and
\begin{equation}
 \frac{{\cal L}_{tot}}{N} \rightarrow \frac{1}{N} S_{config} + S_{vib} \;\sim\; S_{vib}\;\;\;\;\;\;(\tau<\tau_\alpha)
\end{equation}
for large $N$.

On the other hand, if snapshots are spread over $\tau \gg \tau_\alpha$, they will each have different, uncorrelated backbones, and we get:
\begin{equation}
 \frac{{\cal L}_{tot}}{N} \rightarrow S_{config} + S_{vib}\;\;\;\;\;\;(\tau\gg\tau_\alpha)
\end{equation}
since the snapshots are then completely independent.
These two limits will be clearly manifested, as we shall see. We refer to this method of compressing the concatenation of many configurations as the Concatenated Information method, or CCI \footnote{Correlating a patch with the surroundings is a strategy also used in the equilibrium definition of states, see Refs. \cite{van1984statistical,newman2022metastates,holler2020one,read2022complexity}}.


As a proof-of-principle, we will apply these ideas to the t154 lattice glass model \citep{LatticeGlassModelsBiroliMezard,DynamicalHeterogeneityInLatticeGlassModels}, which has been shown to exhibit glassy behavior, avoiding crystallization due to the clashing 'crystallization motifs' of different particle types \cite{DynamicalHeterogeneityInLatticeGlassModels}.
%
%
%
The model employs three types of particles which can occupy the lattice sites of a 3D simple cubic lattice of size $L \times L\times L$. The particle types are labeled by an integer $m$, which can be either 1, 2, or 3. A particle of type $m$ which has $n$ occupied nearest neighbor sites has energy $E_m = max(n-m,0)$. Thus, a particle of type $m$ having $m$ or fewer neighbors has zero energy. The relative abundances of the different particle types are specified to be: 10\% particles of type 1, 50\% type 2, and 40\% type 3 (hence the model's name). The total particle density $\rho$ is a control parameter. The thermodynamics of the system are explored by Monte-Carlo dynamics, employing only simple transitional moves of particles.
 By studying quantities such as average mean squared displacement, $\chi_4$, and the structure factor, Reference \cite{DynamicalHeterogeneityInLatticeGlassModels} confirmed that this model exhibits glassy behavior at high enough densities (Also, see Supplementary Material). Our results are for $L=15$ unless otherwise specified.

Before presenting results on the t154 model, we will consider the simpler example of a 2D Ising model, where part of the system has (artificially) slow dynamics. We start from from an equilibrated state at temperature $T$, choose a fraction $f$ of the spins at random, and give them slow (``frozen'') dynamics, leaving the rest to flip freely. The system is then evolved under Monte-Carlo dynamics at temperature $T$, with a Monte-Carlo flip of the frozen spins accepted with a likelihood which is smaller by a factor $p$ than a similar ``normal'' spin.
The ``frozen'' spins, together with those neighboring spins which are strongly coupled to them, constitute the analog of the backbone in the glassy system.

\begin{figure}[t]
 \centering
 \includegraphics[width=.6\linewidth]{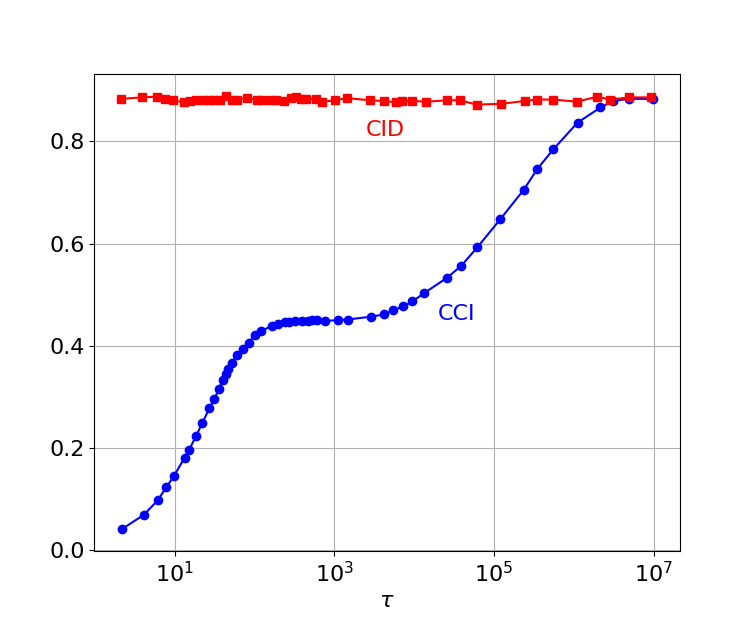}
 \caption{Concatenated and independent information (the latter an estimates of the total entropy) vs. $\tau$ for a 2D Ising system of size $L=50$, with $k_B T=3J$, $f=0.8$, and $p=1\mathrm{e}{-5}$. A single unit of time represents $L^2$ Monte-Carlo steps. The red points are the CID derived from the compression of a single snapshot taken at time $\tau$, and represent the total entropy of the system at that time. Results are averaged over 50 instances of the system. The error on the CCI and CID calculations are on the order of the data points.}
 \label{CCI_Ising}
\end{figure}

 The CCI shown in Fig. \ref{CCI_Ising} exhibits a rapid rise to a plateau at about $\tau \sim 100$ which persists for some time ($\sim 10^4$), eventually rising to a second, higher plateau. By analogy to glasses, we will call the time-scale over which the lower plateau persists $\tau_{\alpha}$. Over this scale, the free spins flip repeatedly, while the frozen spins largely remain unchanged, reminiscent of the backbone of glassy materials. For $\tau<\tau_{\alpha}$, the different snapshots of the system may be thought of as consisting of a fraction $f$ of largely unchanged spins, and a fraction $1-f$ of spins which are largely thermalized, and uncorrelated from snapshot to snapshot \footnote{There are also some free spins which maintain their orientation for over $\tau_\alpha$ since they interact strongly with some frozen spins, for example a single free spin surrounded by frozen spins of the same orientation.}. This partial decoherence is the source of the lower plateau. For $\tau \sim \tau_{\alpha}$, the frozen spins begin to flip as well, leading to a greater loss of correlation between snapshots, with a concomitant rise in the CCI. For $\tau \gg \tau_{\alpha}$, all of the spins in the snapshots are uncorrelated, resulting in the second, higher plateau at $\tau \sim 10^7$. 

As discussed in the Introduction, we identify the height of the lower plateau as the vibrational entropy of the system, and that of the higher plateau as the total entropy of the system. We verify the latter by computing the entropy of the system by compressing single configurations, as was done in Reference \cite{martiniani2019quantifying} where it was called the CID. As can be seen in Fig. \ref{CCI_Ising}, the second plateau is at the value of the CID, consistent with its identification as the total entropy of the system. We conclude that the difference between the heights of the two plateaus is the configurational entropy, that is, the entropy associated with the slow degrees of freedom. As the systems increase in size, these estimates get better and better, converging to the known values in the thermodynamic limit; see the Supplementary Material.



We now consider the CCI of the t154 model. We expect that, as in the Ising system, the glassy system's CCI should reach a plateau quickly, in the time it takes for a robust backbone to form. This backbone, and its associated plateau, should be stable for about $\tau_\alpha$, when the CCI should begin to exit this first plateau and rise to a second, higher plateau. 
The difference between the plateaus gives us the system's configurational entropy directly. 

\begin{figure}[t]
 \centering
 \includegraphics[width=.6\linewidth]{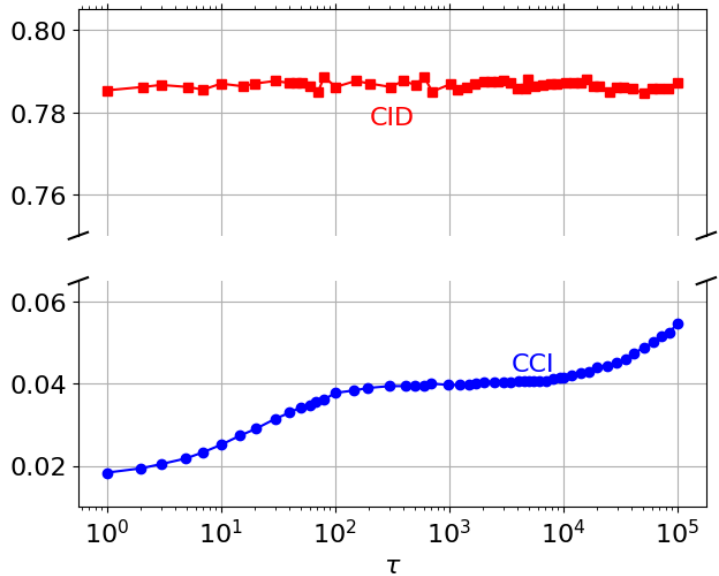}
 \caption{$CCI(\tau)$ and $CID(\tau)$ vs. $\tau$ for the t154 system, with $L=15$, $\rho=0.56$, $T=0.085$, and $t_w=3e5$. A single unit of time represents $L^3$ Monte-Carlo steps. The red points are the CID derived from the compression of a single snapshot taken at time $\tau$, and represent the total entropy of the system at that time. Results are averaged over 50 instances of the system. The error on the CCI and CID calculations are on the order of the data points.}
 \label{CCI vs. t glass example}
\end{figure}

\begin{figure}[]
 \centering
 \includegraphics[width=.6\linewidth]{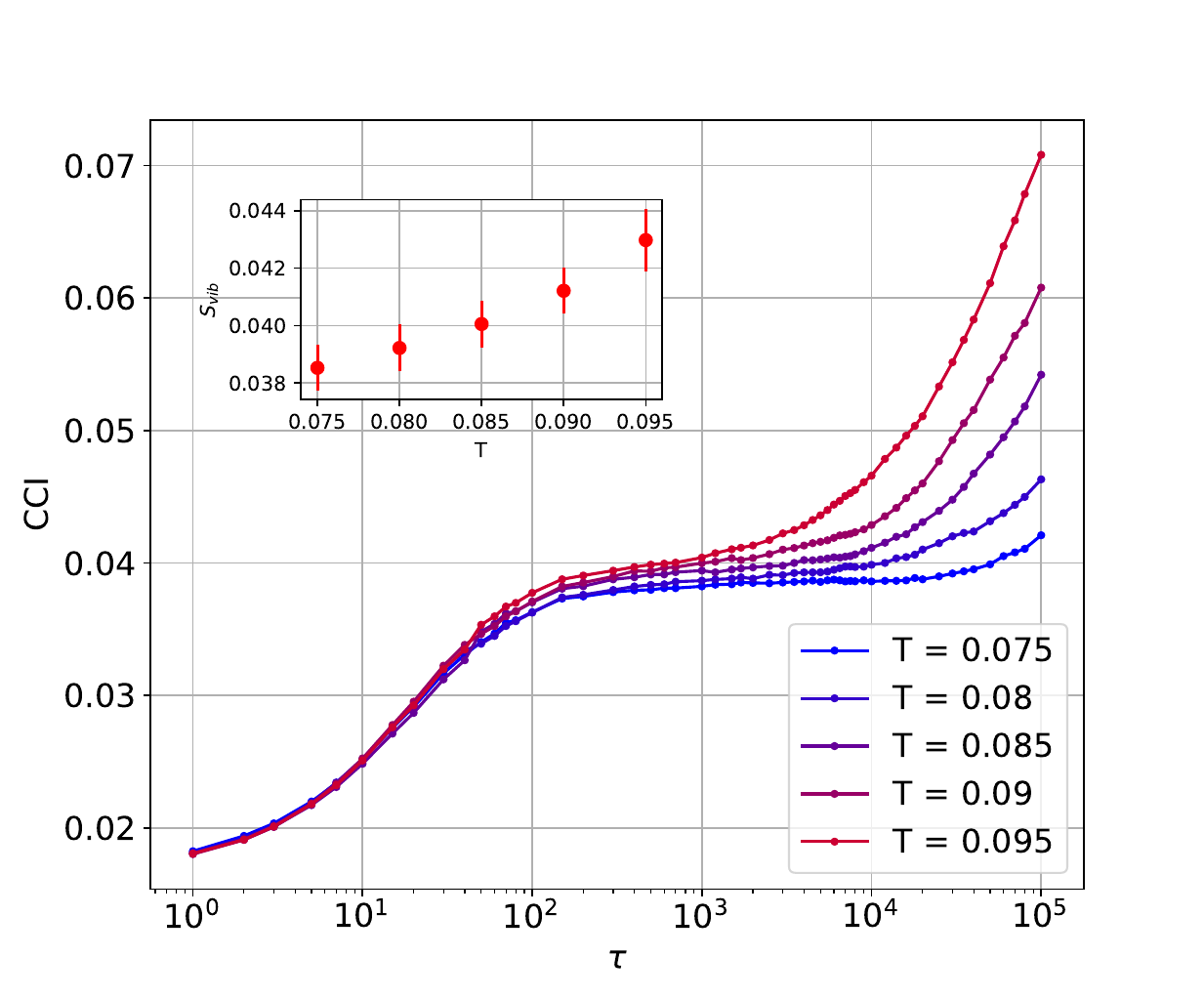}
 \caption{$CCI(\tau)$ vs. $\tau$ for different $T$ for the t154 system, with $L=15, \rho=0.56, t_w=3e5$. A single unit of time represents $L^3$ Monte-Carlo steps. Results are averaged over 50 instances of each system. Inset: Dependence of $S_{vib}$ on $T$.}
 \label{CCI vs. t for different T}
\end{figure}

\begin{figure}[]
\centering
\includegraphics[width=.6\linewidth]{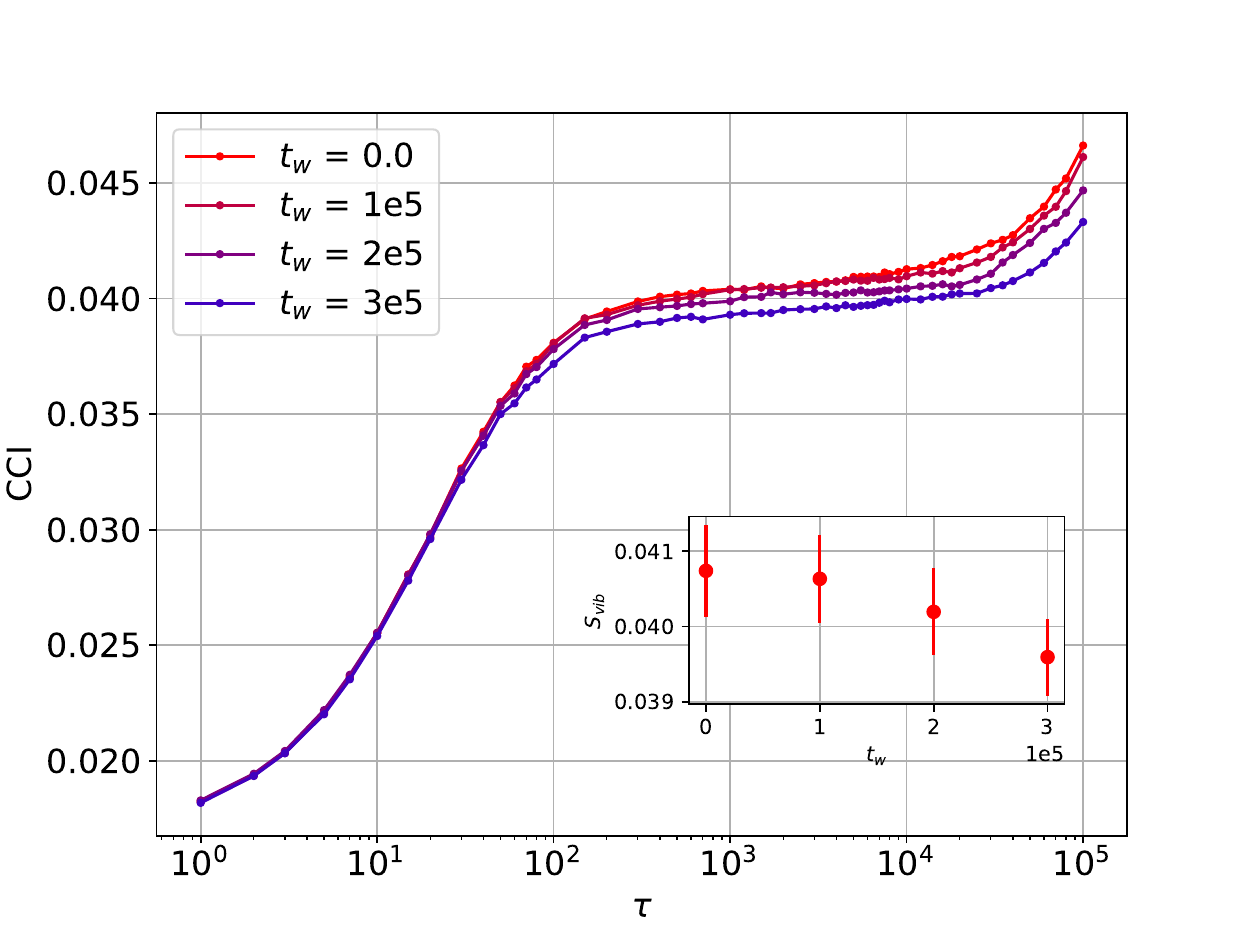}
\caption{$CCI(\tau)$ vs. $\tau$ for different waiting times for the t154 system, with $L=15, \rho=0.56, T=0.075$. A single unit of time represents $L^3$ Monte-Carlo steps. Results are averaged over 50 instances of each system. Inset: Dependence of $S_{vib}$ on waiting time.}
 \label{CCI vs. t for different t_w}
\end{figure}

As is clear from Fig. \ref{CCI vs. t glass example}, the time scale for the rise from the first plateau to the second is very large, and is numerically inaccessible. However, as for the Ising system, by computing the CID of single snapshots at different times, we can measure the total entropy of the system as a function of time, which is shown by the orange symbols. Thus we can estimate the configurational entropy even though we can not follow the full time-evolution of the curve. We may now use these measures to analyze the configurational entropy and $\alpha$-scale of the t154 system as a function of its density, temperature, and aging.

Figures \ref{CCI vs. t for different T} and \ref{CCI vs. t for different t_w} show the $CCI(\tau)$ for different temperatures and waiting times respectively. As can be seen, lower temperatures and longer waiting times result in longer $\alpha$-time scales. The dependence of the vibrational entropy on the temperature and waiting time is small but systematic. As the temperature increases, we expect more random (vibrational) motion around the backbone, which is seen in the inset to Fig. \ref{CCI vs. t for different T}. The dependence on waiting time seen in the inset to Fig. \ref{CCI vs. t for different t_w} is likely due to the system finding a better backbone by finding a deeper energy minimum. This backbone might be better organized, or it could contain more particles, allowing less vibration to occur. 
 
Although the system we study is small, we are able to extract the temperature dependence of $\tau_\alpha$; this is presented in the Supplementary Material. Our measurements show that $\tau_\alpha$ seems to fit the Vogel-Fulcher-Tamman form $\tau_\alpha \sim e^{B/(T-T_0)}$ better than either an Arrhenius or a super-Arrhenius dependence.

Having established that long-lived backbones exist in this model, we now ask what are the building blocks which compose them, and inquire as to their relation to the system's correlation length. In the Supplementary Material we exhibit the 6 most common 2x2x2 patches for a given temperature and waiting time. There are no 3x3x3 repeating patches in our system, a fact that is consistent with a correlation length of $l = 2$ obtained through the decimation method of Reference \cite{martiniani2020correlation}. Other temperatures and waiting times show similar behavior. This is also consistent with our calculation of the point-to-set length for this system, as discussed in the Supplementary Material.

We have argued that for systems with large separation of time scales, the high and low frequency portions can be evaluated independently and directly, without recourse to model-based interpretation. In a sense, the CCI derives naturally from ideas on overlaps in glassy systems which have been studied for many years. However, the fact that it can be directly connected with the total entropy in the large $\tau$ limit facilitates its interpretation in measuring the fast-scale vibrational and slow-scale configurational entropies, which is particularly relevant for glassy systems. 

 We would like to thank Ludovic Berthier, Giulio Biroli, Paul Chaikin, Stefano Martiniani and Misaki Osawa for enlightening discussions. D.~L. thanks the Israel Science Foundation (Grant No. 2083/23). J.~K. is supported by the Simons Foundation Grant. No 454943




\begin{thebibliography}{0}%
\makeatletter
\providecommand \@ifxundefined [1]{%
 \@ifx{#1\undefined}
}%
\providecommand \@ifnum [1]{%
 \ifnum #1\expandafter \@firstoftwo
 \else \expandafter \@secondoftwo
 \fi
}%
\providecommand \@ifx [1]{%
 \ifx #1\expandafter \@firstoftwo
 \else \expandafter \@secondoftwo
 \fi
}%
\providecommand \natexlab [1]{#1}%
\providecommand \enquote  [1]{``#1''}%
\providecommand \bibnamefont  [1]{#1}%
\providecommand \bibfnamefont [1]{#1}%
\providecommand \citenamefont [1]{#1}%
\providecommand \href@noop [0]{\@secondoftwo}%
\providecommand \href [0]{\begingroup \@sanitize@url \@href}%
\providecommand \@href[1]{\@@startlink{#1}\@@href}%
\providecommand \@@href[1]{\endgroup#1\@@endlink}%
\providecommand \@sanitize@url [0]{\catcode `\\12\catcode `\$12\catcode
  `\&12\catcode `\#12\catcode `\^12\catcode `\_12\catcode `\%12\relax}%
\providecommand \@@startlink[1]{}%
\providecommand \@@endlink[0]{}%
\providecommand \url  [0]{\begingroup\@sanitize@url \@url }%
\providecommand \@url [1]{\endgroup\@href {#1}{\urlprefix }}%
\providecommand \urlprefix  [0]{URL }%
\providecommand \Eprint [0]{\href }%
\providecommand \doibase [0]{http://dx.doi.org/}%
\providecommand \selectlanguage [0]{\@gobble}%
\providecommand \bibinfo  [0]{\@secondoftwo}%
\providecommand \bibfield  [0]{\@secondoftwo}%
\providecommand \translation [1]{[#1]}%
\providecommand \BibitemOpen [0]{}%
\providecommand \bibitemStop [0]{}%
\providecommand \bibitemNoStop [0]{.\EOS\space}%
\providecommand \EOS [0]{\spacefactor3000\relax}%
\providecommand \BibitemShut  [1]{\csname bibitem#1\endcsname}%
\let\auto@bib@innerbib\@empty
\end{thebibliography}%


\newpage
\begin{thebibliography}{35}%
\makeatletter
\providecommand \@ifxundefined [1]{%
 \@ifx{#1\undefined}
}%
\providecommand \@ifnum [1]{%
 \ifnum #1\expandafter \@firstoftwo
 \else \expandafter \@secondoftwo
 \fi
}%
\providecommand \@ifx [1]{%
 \ifx #1\expandafter \@firstoftwo
 \else \expandafter \@secondoftwo
 \fi
}%
\providecommand \natexlab [1]{#1}%
\providecommand \enquote  [1]{``#1''}%
\providecommand \bibnamefont  [1]{#1}%
\providecommand \bibfnamefont [1]{#1}%
\providecommand \citenamefont [1]{#1}%
\providecommand \href@noop [0]{\@secondoftwo}%
\providecommand \href [0]{\begingroup \@sanitize@url \@href}%
\providecommand \@href[1]{\@@startlink{#1}\@@href}%
\providecommand \@@href[1]{\endgroup#1\@@endlink}%
\providecommand \@sanitize@url [0]{\catcode `\\12\catcode `\$12\catcode
  `\&12\catcode `\#12\catcode `\^12\catcode `\_12\catcode `\%12\relax}%
\providecommand \@@startlink[1]{}%
\providecommand \@@endlink[0]{}%
\providecommand \url  [0]{\begingroup\@sanitize@url \@url }%
\providecommand \@url [1]{\endgroup\@href {#1}{\urlprefix }}%
\providecommand \urlprefix  [0]{URL }%
\providecommand \Eprint [0]{\href }%
\providecommand \doibase [0]{http://dx.doi.org/}%
\providecommand \selectlanguage [0]{\@gobble}%
\providecommand \bibinfo  [0]{\@secondoftwo}%
\providecommand \bibfield  [0]{\@secondoftwo}%
\providecommand \translation [1]{[#1]}%
\providecommand \BibitemOpen [0]{}%
\providecommand \bibitemStop [0]{}%
\providecommand \bibitemNoStop [0]{.\EOS\space}%
\providecommand \EOS [0]{\spacefactor3000\relax}%
\providecommand \BibitemShut  [1]{\csname bibitem#1\endcsname}%
\let\auto@bib@innerbib\@empty
\bibitem [{\citenamefont {Cavagna}(2009)}]{SuperCooledLiquidsForPedestrians}%
  \BibitemOpen
  \bibfield  {author} {\bibinfo {author} {\bibfnamefont {A.}~\bibnamefont
  {Cavagna}},\ }\href {\doibase :10.1016/j.physrep.2009.03.003} {\bibfield
  {journal} {\bibinfo  {journal} {Physics Reports}\ }\textbf {\bibinfo {volume}
  {476}} (\bibinfo {year} {2009}),\ :10.1016/j.physrep.2009.03.003}\BibitemShut
  {NoStop}%
\bibitem [{\citenamefont {Kauzmann}(1948)}]{kauzmann1948nature}%
  \BibitemOpen
  \bibfield  {author} {\bibinfo {author} {\bibfnamefont {W.}~\bibnamefont
  {Kauzmann}},\ }\href@noop {} {\bibfield  {journal} {\bibinfo  {journal}
  {Chemical reviews}\ }\textbf {\bibinfo {volume} {43}},\ \bibinfo {pages}
  {219} (\bibinfo {year} {1948})}\BibitemShut {NoStop}%
\bibitem [{Note1()}]{Note1}%
  \BibitemOpen
  \bibinfo {note} {Because we shall work in regimes (liquid in equilibrium or
  aging) in which the `backbones' are only temporary, the definition of
  complexity we discuss here is in the thermodynamic limit, inevitably based
  on a timescale, and thus makes sense only in relation to a separation of fast
  (`vibrational') and slow (`alpha') timescales. This is unlike a construction
  applied to the equilibrium Gibbs measure, see References \cite
  {holler2020one,read2022complexity,newman2022metastates}, which apply to
  infinitely long equilibration times.}\BibitemShut {Stop}%
\bibitem [{\citenamefont {Gibbs}\ and\ \citenamefont
  {DiMarzio}(1958)}]{gibbs1958nature}%
  \BibitemOpen
  \bibfield  {author} {\bibinfo {author} {\bibfnamefont {J.~H.}\ \bibnamefont
  {Gibbs}}\ and\ \bibinfo {author} {\bibfnamefont {E.~A.}\ \bibnamefont
  {DiMarzio}},\ }\href@noop {} {\bibfield  {journal} {\bibinfo  {journal} {The
  Journal of Chemical Physics}\ }\textbf {\bibinfo {volume} {28}},\ \bibinfo
  {pages} {373} (\bibinfo {year} {1958})}\BibitemShut {NoStop}%
\bibitem [{\citenamefont {Adam}\ and\ \citenamefont
  {Gibbs}(1965)}]{adam1965temperature}%
  \BibitemOpen
  \bibfield  {author} {\bibinfo {author} {\bibfnamefont {G.}~\bibnamefont
  {Adam}}\ and\ \bibinfo {author} {\bibfnamefont {J.~H.}\ \bibnamefont
  {Gibbs}},\ }\href@noop {} {\bibfield  {journal} {\bibinfo  {journal} {The
  journal of chemical physics}\ }\textbf {\bibinfo {volume} {43}},\ \bibinfo
  {pages} {139} (\bibinfo {year} {1965})}\BibitemShut {NoStop}%
\bibitem [{\citenamefont {Kurchan}\ and\ \citenamefont
  {Levine}(2010)}]{kurchan2010order}%
  \BibitemOpen
  \bibfield  {author} {\bibinfo {author} {\bibfnamefont {J.}~\bibnamefont
  {Kurchan}}\ and\ \bibinfo {author} {\bibfnamefont {D.}~\bibnamefont
  {Levine}},\ }\href@noop {} {\bibfield  {journal} {\bibinfo  {journal}
  {Journal of Physics A: Mathematical and Theoretical}\ }\textbf {\bibinfo
  {volume} {44}},\ \bibinfo {pages} {035001} (\bibinfo {year}
  {2010})}\BibitemShut {NoStop}%
\bibitem [{\citenamefont {Berthier}\ \emph {et~al.}(2011)\citenamefont
  {Berthier}, \citenamefont {Biroli}, \citenamefont {Bouchaud},\ and\
  \citenamefont {Jack}}]{berthier2011overview}%
  \BibitemOpen
  \bibfield  {author} {\bibinfo {author} {\bibfnamefont {L.}~\bibnamefont
  {Berthier}}, \bibinfo {author} {\bibfnamefont {G.}~\bibnamefont {Biroli}},
  \bibinfo {author} {\bibfnamefont {J.-P.}\ \bibnamefont {Bouchaud}}, \ and\
  \bibinfo {author} {\bibfnamefont {R.~L.}\ \bibnamefont {Jack}},\ }\href@noop
  {} {\bibfield  {journal} {\bibinfo  {journal} {Dynamical Heterogeneities in
  Glasses, Colloids, and Granular Media}\ }\textbf {\bibinfo {volume} {150}},\
  \bibinfo {pages} {68} (\bibinfo {year} {2011})}\BibitemShut {NoStop}%
\bibitem [{\citenamefont {Van~Enter}\ and\ \citenamefont {van
  Hemmen}(1984)}]{van1984statistical}%
  \BibitemOpen
  \bibfield  {author} {\bibinfo {author} {\bibfnamefont {A.}~\bibnamefont
  {Van~Enter}}\ and\ \bibinfo {author} {\bibfnamefont {J.~L.}\ \bibnamefont
  {van Hemmen}},\ }\href@noop {} {\bibfield  {journal} {\bibinfo  {journal}
  {Physical Review A}\ }\textbf {\bibinfo {volume} {29}},\ \bibinfo {pages}
  {355} (\bibinfo {year} {1984})}\BibitemShut {NoStop}%
\bibitem [{\citenamefont {Read}(2022)}]{read2022complexity}%
  \BibitemOpen
  \bibfield  {author} {\bibinfo {author} {\bibfnamefont {N.}~\bibnamefont
  {Read}},\ }\href@noop {} {\bibfield  {journal} {\bibinfo  {journal} {Physical
  Review E}\ }\textbf {\bibinfo {volume} {105}},\ \bibinfo {pages} {054134}
  (\bibinfo {year} {2022})}\BibitemShut {NoStop}%
\bibitem [{\citenamefont {Berthier}\ \emph
  {et~al.}(2017{\natexlab{a}})\citenamefont {Berthier}, \citenamefont
  {Charbonneau}, \citenamefont {Coslovich}, \citenamefont {Ninarello},
  \citenamefont {Ozawa},\ and\ \citenamefont
  {Yaida}}]{berthier2017configurational}%
  \BibitemOpen
  \bibfield  {author} {\bibinfo {author} {\bibfnamefont {L.}~\bibnamefont
  {Berthier}}, \bibinfo {author} {\bibfnamefont {P.}~\bibnamefont
  {Charbonneau}}, \bibinfo {author} {\bibfnamefont {D.}~\bibnamefont
  {Coslovich}}, \bibinfo {author} {\bibfnamefont {A.}~\bibnamefont
  {Ninarello}}, \bibinfo {author} {\bibfnamefont {M.}~\bibnamefont {Ozawa}}, \
  and\ \bibinfo {author} {\bibfnamefont {S.}~\bibnamefont {Yaida}},\
  }\href@noop {} {\bibfield  {journal} {\bibinfo  {journal} {Proceedings of the
  National Academy of Sciences}\ }\textbf {\bibinfo {volume} {114}},\ \bibinfo
  {pages} {11356} (\bibinfo {year} {2017}{\natexlab{a}})}\BibitemShut {NoStop}%
\bibitem [{\citenamefont {Frenkel}\ and\ \citenamefont
  {Ladd}(1984)}]{frenkel1984new}%
  \BibitemOpen
  \bibfield  {author} {\bibinfo {author} {\bibfnamefont {D.}~\bibnamefont
  {Frenkel}}\ and\ \bibinfo {author} {\bibfnamefont {A.~J.}\ \bibnamefont
  {Ladd}},\ }\href@noop {} {\bibfield  {journal} {\bibinfo  {journal} {The
  Journal of chemical physics}\ }\textbf {\bibinfo {volume} {81}},\ \bibinfo
  {pages} {3188} (\bibinfo {year} {1984})}\BibitemShut {NoStop}%
\bibitem [{\citenamefont {Franz}\ and\ \citenamefont
  {Parisi}(1998)}]{franz1998effective}%
  \BibitemOpen
  \bibfield  {author} {\bibinfo {author} {\bibfnamefont {S.}~\bibnamefont
  {Franz}}\ and\ \bibinfo {author} {\bibfnamefont {G.}~\bibnamefont {Parisi}},\
  }\href@noop {} {\bibfield  {journal} {\bibinfo  {journal} {Physica A:
  Statistical Mechanics and its Applications}\ }\textbf {\bibinfo {volume}
  {261}},\ \bibinfo {pages} {317} (\bibinfo {year} {1998})}\BibitemShut
  {NoStop}%
\bibitem [{Note2()}]{Note2}%
  \BibitemOpen
  \bibinfo {note} {The Franz-Parisi method relies on the reference system being
  in equilibrium, otherwise the connection to the relevant theory is
  lost.}\BibitemShut {Stop}%
\bibitem [{\citenamefont {Kirkpatrick}\ and\ \citenamefont
  {Thirumalai}(1987)}]{kirkpatrick1987p}%
  \BibitemOpen
  \bibfield  {author} {\bibinfo {author} {\bibfnamefont {T.~R.}\ \bibnamefont
  {Kirkpatrick}}\ and\ \bibinfo {author} {\bibfnamefont {D.}~\bibnamefont
  {Thirumalai}},\ }\href@noop {} {\bibfield  {journal} {\bibinfo  {journal}
  {Physical Review B}\ }\textbf {\bibinfo {volume} {36}},\ \bibinfo {pages}
  {5388} (\bibinfo {year} {1987})}\BibitemShut {NoStop}%
\bibitem [{\citenamefont {Kirkpatrick}\ \emph {et~al.}(1989)\citenamefont
  {Kirkpatrick}, \citenamefont {Thirumalai},\ and\ \citenamefont
  {Wolynes}}]{kirkpatrick1989scaling}%
  \BibitemOpen
  \bibfield  {author} {\bibinfo {author} {\bibfnamefont {T.~R.}\ \bibnamefont
  {Kirkpatrick}}, \bibinfo {author} {\bibfnamefont {D.}~\bibnamefont
  {Thirumalai}}, \ and\ \bibinfo {author} {\bibfnamefont {P.~G.}\ \bibnamefont
  {Wolynes}},\ }\href@noop {} {\bibfield  {journal} {\bibinfo  {journal}
  {Physical Review A}\ }\textbf {\bibinfo {volume} {40}},\ \bibinfo {pages}
  {1045} (\bibinfo {year} {1989})}\BibitemShut {NoStop}%
\bibitem [{\citenamefont {Lubchenko}\ and\ \citenamefont
  {Wolynes}(2006)}]{lubchenko2006theory}%
  \BibitemOpen
  \bibfield  {author} {\bibinfo {author} {\bibfnamefont {V.}~\bibnamefont
  {Lubchenko}}\ and\ \bibinfo {author} {\bibfnamefont {P.~G.}\ \bibnamefont
  {Wolynes}},\ }\href@noop {} {\bibfield  {journal} {\bibinfo  {journal} {arXiv
  preprint cond-mat/0607349}\ } (\bibinfo {year} {2006})}\BibitemShut {NoStop}%
\bibitem [{\citenamefont {Berthier}\ \emph
  {et~al.}(2017{\natexlab{b}})\citenamefont {Berthier}, \citenamefont
  {Charbonneau}, \citenamefont {Coslovich}, \citenamefont {Ninarello},
  \citenamefont {Ozawa},\ and\ \citenamefont
  {Yaida}}]{ConfigEntropyCalcBerthier}%
  \BibitemOpen
  \bibfield  {author} {\bibinfo {author} {\bibfnamefont {L.}~\bibnamefont
  {Berthier}}, \bibinfo {author} {\bibfnamefont {P.}~\bibnamefont
  {Charbonneau}}, \bibinfo {author} {\bibfnamefont {D.}~\bibnamefont
  {Coslovich}}, \bibinfo {author} {\bibfnamefont {A.}~\bibnamefont
  {Ninarello}}, \bibinfo {author} {\bibfnamefont {M.}~\bibnamefont {Ozawa}}, \
  and\ \bibinfo {author} {\bibfnamefont {S.}~\bibnamefont {Yaida}},\ }\href
  {\doibase 10.1073/pnas.1706860114} {\bibfield  {journal} {\bibinfo  {journal}
  {Proceedings of the National Academy of Sciences}\ }\textbf {\bibinfo
  {volume} {114}},\ \bibinfo {pages} {11356} (\bibinfo {year}
  {2017}{\natexlab{b}})},\ \Eprint
  {http://arxiv.org/abs/https://www.pnas.org/content/114/43/11356.full.pdf}
  {https://www.pnas.org/content/114/43/11356.full.pdf} \BibitemShut {NoStop}%
\bibitem [{\citenamefont {Bouchaud}\ and\ \citenamefont
  {Biroli}(2004)}]{bouchaud2004adam}%
  \BibitemOpen
  \bibfield  {author} {\bibinfo {author} {\bibfnamefont {J.-P.}\ \bibnamefont
  {Bouchaud}}\ and\ \bibinfo {author} {\bibfnamefont {G.}~\bibnamefont
  {Biroli}},\ }\href@noop {} {\bibfield  {journal} {\bibinfo  {journal} {The
  Journal of chemical physics}\ }\textbf {\bibinfo {volume} {121}},\ \bibinfo
  {pages} {7347} (\bibinfo {year} {2004})}\BibitemShut {NoStop}%
\bibitem [{\citenamefont {Cover}\ and\ \citenamefont
  {Thomas}(2012)}]{cover2012elements}%
  \BibitemOpen
  \bibfield  {author} {\bibinfo {author} {\bibfnamefont {T.~M.}\ \bibnamefont
  {Cover}}\ and\ \bibinfo {author} {\bibfnamefont {J.~A.}\ \bibnamefont
  {Thomas}},\ }\href@noop {} {\emph {\bibinfo {title} {Elements of information
  theory}}}\ (\bibinfo  {publisher} {John Wiley \& Sons},\ \bibinfo {year}
  {2012})\BibitemShut {NoStop}%
\bibitem [{\citenamefont {Ziv}\ and\ \citenamefont
  {Lempel}(1977)}]{ziv1977universal}%
  \BibitemOpen
  \bibfield  {author} {\bibinfo {author} {\bibfnamefont {J.}~\bibnamefont
  {Ziv}}\ and\ \bibinfo {author} {\bibfnamefont {A.}~\bibnamefont {Lempel}},\
  }\href@noop {} {\bibfield  {journal} {\bibinfo  {journal} {IEEE Transactions
  on Information Theory}\ }\textbf {\bibinfo {volume} {23}},\ \bibinfo {pages}
  {337} (\bibinfo {year} {1977})}\BibitemShut {NoStop}%
\bibitem [{\citenamefont {Ziv}\ and\ \citenamefont
  {Lempel}(1978)}]{ziv1978compression}%
  \BibitemOpen
  \bibfield  {author} {\bibinfo {author} {\bibfnamefont {J.}~\bibnamefont
  {Ziv}}\ and\ \bibinfo {author} {\bibfnamefont {A.}~\bibnamefont {Lempel}},\
  }\href@noop {} {\bibfield  {journal} {\bibinfo  {journal} {IEEE Transactions
  on Information Theory}\ }\textbf {\bibinfo {volume} {24}},\ \bibinfo {pages}
  {530} (\bibinfo {year} {1978})}\BibitemShut {NoStop}%
\bibitem [{\citenamefont {Shannon}(1948)}]{shannon2001mathematical}%
  \BibitemOpen
  \bibfield  {author} {\bibinfo {author} {\bibfnamefont {C.~E.}\ \bibnamefont
  {Shannon}},\ }\href@noop {} {\bibfield  {journal} {\bibinfo  {journal} {Bell
  System Technial Journal}\ }\textbf {\bibinfo {volume} {27}},\ \bibinfo
  {pages} {379} (\bibinfo {year} {1948})}\BibitemShut {NoStop}%
\bibitem [{Note3()}]{Note3}%
  \BibitemOpen
  \bibinfo {note} {As with any classical entropy, one has to identify all
  continuous quantities up to an error $\epsilon $: densities, and, when
  necessary, positions, radii and eccentricities of particles. The results then
  scale logarithmically with the coarse-graining scale $\epsilon $ which drops
  off in the thermodynamic limit.}\BibitemShut {Stop}%
\bibitem [{\citenamefont {Martiniani}\ \emph {et~al.}(2019)\citenamefont
  {Martiniani}, \citenamefont {Chaikin},\ and\ \citenamefont
  {Levine}}]{martiniani2019quantifying}%
  \BibitemOpen
  \bibfield  {author} {\bibinfo {author} {\bibfnamefont {S.}~\bibnamefont
  {Martiniani}}, \bibinfo {author} {\bibfnamefont {P.~M.}\ \bibnamefont
  {Chaikin}}, \ and\ \bibinfo {author} {\bibfnamefont {D.}~\bibnamefont
  {Levine}},\ }\href@noop {} {\bibfield  {journal} {\bibinfo  {journal}
  {Physical Review X}\ }\textbf {\bibinfo {volume} {9}},\ \bibinfo {pages}
  {011031} (\bibinfo {year} {2019})}\BibitemShut {NoStop}%
\bibitem [{\citenamefont {Avinery}\ \emph {et~al.}(2019)\citenamefont
  {Avinery}, \citenamefont {Kornreich},\ and\ \citenamefont
  {Beck}}]{avinery2019universal}%
  \BibitemOpen
  \bibfield  {author} {\bibinfo {author} {\bibfnamefont {R.}~\bibnamefont
  {Avinery}}, \bibinfo {author} {\bibfnamefont {M.}~\bibnamefont {Kornreich}},
  \ and\ \bibinfo {author} {\bibfnamefont {R.}~\bibnamefont {Beck}},\
  }\href@noop {} {\bibfield  {journal} {\bibinfo  {journal} {Physical Review
  Letters}\ }\textbf {\bibinfo {volume} {123}},\ \bibinfo {pages} {178102}
  (\bibinfo {year} {2019})}\BibitemShut {NoStop}%
\bibitem [{\citenamefont {Martiniani}\ \emph {et~al.}(2020)\citenamefont
  {Martiniani}, \citenamefont {Lemberg}, \citenamefont {Chaikin},\ and\
  \citenamefont {Levine}}]{martiniani2020correlation}%
  \BibitemOpen
  \bibfield  {author} {\bibinfo {author} {\bibfnamefont {S.}~\bibnamefont
  {Martiniani}}, \bibinfo {author} {\bibfnamefont {Y.}~\bibnamefont {Lemberg}},
  \bibinfo {author} {\bibfnamefont {P.~M.}\ \bibnamefont {Chaikin}}, \ and\
  \bibinfo {author} {\bibfnamefont {D.}~\bibnamefont {Levine}},\ }\href@noop {}
  {\bibfield  {journal} {\bibinfo  {journal} {Physical Review Letters}\
  }\textbf {\bibinfo {volume} {125}},\ \bibinfo {pages} {170601} (\bibinfo
  {year} {2020})}\BibitemShut {NoStop}%
\bibitem [{\citenamefont {Cavagna}\ \emph {et~al.}(2020)\citenamefont
  {Cavagna}, \citenamefont {Chaikin}, \citenamefont {Levine}, \citenamefont
  {Martiniani}, \citenamefont {Puglisi},\ and\ \citenamefont
  {Viale}}]{cavagna2020vicsek}%
  \BibitemOpen
  \bibfield  {author} {\bibinfo {author} {\bibfnamefont {A.}~\bibnamefont
  {Cavagna}}, \bibinfo {author} {\bibfnamefont {P.~M.}\ \bibnamefont
  {Chaikin}}, \bibinfo {author} {\bibfnamefont {D.}~\bibnamefont {Levine}},
  \bibinfo {author} {\bibfnamefont {S.}~\bibnamefont {Martiniani}}, \bibinfo
  {author} {\bibfnamefont {A.}~\bibnamefont {Puglisi}}, \ and\ \bibinfo
  {author} {\bibfnamefont {M.}~\bibnamefont {Viale}},\ }\href@noop {}
  {\bibfield  {journal} {\bibinfo  {journal} {Physical Review E}\
  } \textbf {\bibinfo {volume} {103}},\ \bibinfo {pages} {062141 }(\bibinfo {year} {2021})}\BibitemShut {NoStop}%
\bibitem [{\citenamefont {Vicsek}\ \emph {et~al.}(1995)\citenamefont {Vicsek},
  \citenamefont {Czir{\'o}k}, \citenamefont {Ben-Jacob}, \citenamefont
  {Cohen},\ and\ \citenamefont {Shochet}}]{vicsek1995novel}%
  \BibitemOpen
  \bibfield  {author} {\bibinfo {author} {\bibfnamefont {T.}~\bibnamefont
  {Vicsek}}, \bibinfo {author} {\bibfnamefont {A.}~\bibnamefont {Czir{\'o}k}},
  \bibinfo {author} {\bibfnamefont {E.}~\bibnamefont {Ben-Jacob}}, \bibinfo
  {author} {\bibfnamefont {I.}~\bibnamefont {Cohen}}, \ and\ \bibinfo {author}
  {\bibfnamefont {O.}~\bibnamefont {Shochet}},\ }\href@noop {} {\bibfield
  {journal} {\bibinfo  {journal} {Physical Review Letters}\ }\textbf {\bibinfo
  {volume} {75}},\ \bibinfo {pages} {1226} (\bibinfo {year}
  {1995})}\BibitemShut {NoStop}%
\bibitem [{\citenamefont {Ro}\ \emph {et~al.}(2022)\citenamefont {Ro},
  \citenamefont {Guo}, \citenamefont {Shih}, \citenamefont {Phan},
  \citenamefont {Austin}, \citenamefont {Levine}, \citenamefont {Chaikin},\
  and\ \citenamefont {Martiniani}}]{ro2022model}%
  \BibitemOpen
  \bibfield  {author} {\bibinfo {author} {\bibfnamefont {S.}~\bibnamefont
  {Ro}}, \bibinfo {author} {\bibfnamefont {B.}~\bibnamefont {Guo}}, \bibinfo
  {author} {\bibfnamefont {A.}~\bibnamefont {Shih}}, \bibinfo {author}
  {\bibfnamefont {T.~V.}\ \bibnamefont {Phan}}, \bibinfo {author}
  {\bibfnamefont {R.~H.}\ \bibnamefont {Austin}}, \bibinfo {author}
  {\bibfnamefont {D.}~\bibnamefont {Levine}}, \bibinfo {author} {\bibfnamefont
  {P.~M.}\ \bibnamefont {Chaikin}}, \ and\ \bibinfo {author} {\bibfnamefont
  {S.}~\bibnamefont {Martiniani}},\ }\href@noop {} {\bibfield  {journal}
  {\bibinfo  {journal} {Physical Review Letters}\ }\textbf {\bibinfo {volume}
  {129}},\ \bibinfo {pages} {220601} (\bibinfo {year} {2022})}\BibitemShut
  {NoStop}%
\bibitem [{Note4()}]{Note4}%
  \BibitemOpen
  \bibinfo {note} {Correlating a patch with the surroundings is a strategy also
  used in the equilibrium definition of states, see Refs. \cite
  {van1984statistical,newman2022metastates,holler2020one,read2022complexity}}\BibitemShut
  {NoStop}%
\bibitem [{\citenamefont {Biroli}\ and\ \citenamefont
  {Mézard}(2002)}]{LatticeGlassModelsBiroliMezard}%
  \BibitemOpen
  \bibfield  {author} {\bibinfo {author} {\bibfnamefont {G.}~\bibnamefont
  {Biroli}}\ and\ \bibinfo {author} {\bibfnamefont {M.}~\bibnamefont
  {Mézard}},\ }\href {\doibase 10.1103/PhysRevLett.88.025501} {\bibfield
  {journal} {\bibinfo  {journal} {Phys. Rev. Lett.}\ }\textbf {\bibinfo
  {volume} {88}} (\bibinfo {year} {2002}),\
  10.1103/PhysRevLett.88.025501}\BibitemShut {NoStop}%
\bibitem [{\citenamefont {R.~Darst}\ and\ \citenamefont
  {Biroli}(2010)}]{DynamicalHeterogeneityInLatticeGlassModels}%
  \BibitemOpen
  \bibfield  {author} {\bibinfo {author} {\bibfnamefont {D.~R.}\ \bibnamefont
  {R.~Darst}}\ and\ \bibinfo {author} {\bibfnamefont {G.}~\bibnamefont
  {Biroli}},\ }\href {\doibase :10.1063/1.3298877} {\bibfield  {journal}
  {\bibinfo  {journal} {J. Chem. Phys.}\ }\textbf {\bibinfo {volume} {132}}
  (\bibinfo {year} {2010}),\ :10.1063/1.3298877}\BibitemShut {NoStop}%
\bibitem [{Note5()}]{Note5}%
  \BibitemOpen
  \bibinfo {note} {There are also some free spins which maintain their
  orientation for over $\tau _\alpha $ since they interact strongly with some
  frozen spins, for example a single free spin surrounded by frozen spins of
  the same orientation.}\BibitemShut {Stop}%
\bibitem [{\citenamefont {H{\"o}ller}\ and\ \citenamefont
  {Read}(2020)}]{holler2020one}%
  \BibitemOpen
  \bibfield  {author} {\bibinfo {author} {\bibfnamefont {J.}~\bibnamefont
  {H{\"o}ller}}\ and\ \bibinfo {author} {\bibfnamefont {N.}~\bibnamefont
  {Read}},\ }\href@noop {} {\bibfield  {journal} {\bibinfo  {journal} {Physical
  Review E}\ }\textbf {\bibinfo {volume} {101}},\ \bibinfo {pages} {042114}
  (\bibinfo {year} {2020})}\BibitemShut {NoStop}%
 %
  \bibitem [{\citenamefont {Newman}\ \emph {et~al.}(2022)\citenamefont {Newman},
   \citenamefont {Read},\ and\ \citenamefont {Stein}}]{newman2022metastates}%
  \BibitemOpen
  \bibfield  {author} {\bibinfo {author} {\bibfnamefont {C.}~\bibnamefont
  {Newman}}, \bibinfo {author} {\bibfnamefont {N.}~\bibnamefont {Read}}, \and\
  \bibinfo {author} {\bibfnamefont {D.}~\bibnamefont {Stein}},\ }\href@noop {}
  {\bibfield  {journal} {\bibinfo  {journal} {Spin Glass Theory and Far Beyond: Replica Symmetry Breaking After 40 Years, 697-718}\
  } (\bibinfo {year} {2023})}\BibitemShut {NoStop}%

  
\end{thebibliography}
%

\end{document}